\documentclass[11pt]{article}
\usepackage{hyperref}
\usepackage{footnote}
\makesavenoteenv{minipage}
\renewcommand{\title}[1]{%
    \bigskip%
    \begin{center}%
    \Large\bf #1%
    \end{center}%
    \vskip .2in}

\renewcommand{\author}[1]{%
    {\begin{center}
    #1
    \end{center}}}
\newcommand{\address}[1]{\vspace{-1.7em}\vspace{0pt}
    {\begin{center}
    \it #1
    \end{center}}}
    
\begin{document}

\begin{titlepage}
\title{\bf New (Ghost-Free) Formulation of the Pais-Uhlenbeck Oscillator }

\author
{
\bf Rabin Banerjee
\\[0.3cm]
\address{S. N. Bose National Centre for Basic Sciences,

 JD Block, Sector III, Salt Lake City, Kolkata -700 098, India }
}

\address{\tt rabin@bose.res.in}
\begin{abstract}
We provide a new formulation of the Pais-Uhlenbeck oscillator which is a prototype of a higher derivative model. Different parametrisations that reveal the model as a combination of two simple harmonic oscillators are introduced. Conventional results are reproduced in one realisation. In another, all problems related to lack of unitarity or boundedness of energy are eliminated since the hamiltonian is expressed as a sum of the hamiltonians of two decoupled harmonic oscillators. Recourse to imaginary scaling transformation or PT-symmetry, as advocated in the literature, are totally avoided.     
\end{abstract}
\end{titlepage} 

It is a well known truism that theories whose dynamics is governed by higher (than second) order equations of motion are problematic because they have states, known as ghosts, with nonpositive norm. The general ideas have been concretised in the context of the fourth order derivative Pais-Uhlenbeck (PU) oscillator model\cite{PU:1950}, the prototype of a higher derivative quantum field theory. This model has been studied extensively [\cite{MD:2005}-\cite{MP:2013}] and possible solutions to the problem of ghosts have been suggested [\cite{BM:2008}, \cite{MOS:2011}, \cite{CFLT:2013}]. However, as we briefly review below, a completely clinching resolution within the framework of standard quantum field theory, is lacking. The purpose of this letter is to fill this gap.

We devolop a new approach to solve the PU model. The solution is presented in different parametrisations. In one case it reproduces the known results in the literature [\cite{MD:2005} - \cite{MOS:2011}]. In another parametrisation all problems associated with the PU model, like appearance of ghosts, lack of unitarity, instability due to the absence of a well defined ground state etc. are removed at one stroke. Indeed we show that the hamiltonian is represented as a sum of the hamiltonians of two simple harmonic oscillators. Consequently within the ambit of standard field theory all problems get resolved.

The lagrangian of the PU model is given by, 
\begin{equation}
L=\frac{1}{2}[\ddot{z}^2-(\omega_1^2+\omega_2^2)\dot{z}^2+\omega_1^2\omega_2^2z^2]
\label{PUlag}
\end{equation} 
where $\omega_1$ and $\omega_2$ are positive constants and without loss of generality,   $\omega_1>\omega_2$. The case $\omega_1=\omega_2$ is singular \cite{MD:2005}. Following the Ostrogradsky approach or alternately, introducing an additional variable via a Lagrange multiplier to account for the higher derivative nature, one obtains the hamiltonian \cite{MD:2005}
\begin{equation}
H=\frac{p_x^2}{2}+p_zx+\frac{1}{2}(\omega_1^2+\omega_2^2)x^2-\frac{1}{2}\omega_1^2\omega_2^2z^2
\label{PUH}
\end{equation}
where the independent canonical pairs are given by $(x,p_x)$ and  $(z,p_z)$. Note that, in terms of the original variables, $x$ has to be identified with $\dot{z}$. The fourth order equation of motion following from either the lagrangian (\ref{PUlag}) or the hamiltonian (\ref{PUH}) is,
\begin{equation}
{z}^{(4)}+(\omega_1^2+\omega_2^2){z}^{(2)}+\omega_1^2\omega_2^2z=0
\label{EOM}
\end{equation}
where $z^{(k)}$ indicates the k$^{th}$ derivative of the real dynamical variable z.

The hamiltonian (\ref{PUH}) is not positive definite leading to various difficulties. These are best encapsulated by diagonalising the hamiltonian using the canonical transformation [\cite{SMILGA:2006}, \cite{MOS:2011}]
\begin{equation}
x_1=\frac{p_z+\omega_1^2x}{\omega_1\sqrt{\omega_1^2-\omega_2^2}},~~~~~~~p_1=\frac{\omega_1(p_x+\omega_2^2z)}{\sqrt{\omega_1^2-\omega_2^2}},
~~~~~~~x_2=\frac{p_x+\omega_1^2z}{\sqrt{\omega_1^2-\omega_2^2}},~~~~~~~~~p_2=\frac{p_z+\omega_2^2x}{\sqrt{\omega_1^2-\omega_2^2}}
\label{Diagct}
\end{equation}
In terms of the new canonical pairs $(x_1,p_1)$ and  $(x_2,p_2)$ the hamiltonian (\ref{PUH}) diagonalises as,
\begin{equation}
H=\frac{1}{2}(p_1^2+\omega_1^2x_1^2)-\frac{1}{2}(p_2^2+\omega_2^2x_2^2)
\label{Hdiff}
\end{equation}

The usual canonical quantization of (\ref{Hdiff}) is straightforward and yields the hermitian hamiltonian operator,
\begin{equation}
\hat{H}=\frac{1}{2}(\hat{p}_1^2+\omega_1^2\hat{x}_1^2)-\frac{1}{2}(\hat{p}_2^2+\omega_2^2\hat{x}_2^2)
\end{equation}
with $\hat{p}_i=-i\frac{\partial}{\partial x_i}, ~~~~(i=1,2)$. The energy eigenvalues are given by $E_{(n_1,n_2)}=\hbar[\omega_1(n_1+1/2)-\omega_2(n_2+1/2)]$
where $n_1,n_2=0,1,2,....$. The quantum system is unstable since the energy is unbounded both from above and below. This may be bypassed by using an indefinite metric quantisation. Although this ensures stability of the quantum theory, it suffers from the appearance of ghost states and hence, a lack of unitarity.

A few years back, Bender and Mannheim \cite{BM:2008} gave a new realization of the PU model that does not have negative norm states and also has a bounded real energy spectrum. Here the hamiltonian is no longer hermitian but it is PT symmetric (i.e. symmetric under combined space reflection $(P)$ and time reversal $(T)$). The usual inner product has to be modified so that ghosts are interpreted as usual states with positive norm. Subsequently this entire construction was shown to be equivalent to the imaginary scaling $x_2\rightarrow -ix_2$ and $p_2\rightarrow ip_2$ \cite{MOS:2011}. This (complex) canonical transformation, when exploited in (\ref{Hdiff}), flips the sign of the second piece. The expression then resembles the sum of the hamiltonians of two harmonic oscillators. The corresponding quantum theory, therefore, may be easily formulated in a consistent matter. However this complex scaling conflicts with the correspondence principle since the classical limit of the ensuing quantum theory does not coincide with the classical PU oscillator [\cite{BM:2008}, \cite{MOS:2011}]. A way out of this impasse was suggested in \cite{MOS:2011} where the quantum analogue of the imaginary scaling transformation was used. However this has its own problems and unusual properties, not the least among them being a change of sign in the creation operator of the theory. It thus appears that the various remedial solutions have their own merits and/or demerits and a full proof resolution within the domain of standard quantum field theory is lacking.

Let us now commence our analysis. It is clear that the PU model represents two oscillators (with frequencies $\omega_1$ and $\omega_2$) coupled by the fourth order equation (\ref{EOM}) as,
\begin{equation}
(\frac{d^2}{dt^2}+\omega_1^2)(\frac{d^2}{dt^2}+\omega_2^2)z=0
\label{Prodos}
\end{equation}

The factorisation of the PU model in terms of a coupled set of standard harmonic oscillators provides the genesis of our formalism. To probe this factorisation consider the lagrangians $(L_1,L_2)$ for two independent harmonic oscillators with frequencies $\omega_1$ and $\omega_2$,
\begin{equation}
L_i=\frac{1}{2}\dot{x_i}^2-\frac{1}{2}\omega_i^2{x_i}^2 ~~~~ (i=1,2)
\label{Sumos}
\end{equation}
where no summation is implied over the index `$i$'. The equations of motion are given by,
\begin{equation}
\ddot{x_i}=-\omega_i^2x_i
\end{equation}
In terms of the new variables,
\begin{equation}
x_1 \pm x_2=z_{\pm}
\label{NVar}
\end{equation}
the equations of motion are expressed as,
\begin{equation}
\ddot{z}_{+}=-\frac{1}{2}(\omega_1^2+\omega_2^2)z_{+}-\frac{1}{2}(\omega_1^2-\omega_2^2)z_{-},
~~~~\ddot{z}_{-}=-\frac{1}{2}(\omega_1^2+\omega_2^2)z_{-}-\frac{1}{2}(\omega_1^2-\omega_2^2)z_{+}
\label{NVeq}
\end{equation}
obtained by adding and subtracting the equations of motion for $x_1$ and $x_2$. The above pair of coupled second order differential equations may be expressed in terms of the decoupled fourth order equation,
\begin{equation}
{z_{\pm}}^{(4)}+(\omega_1^2+\omega_2^2){z_{\pm}}^{(2)}+\omega_1^2\omega_2^2z_{\pm}=0
\end{equation}
This reproduces the PU equation of motion (\ref{EOM}) for both $z_{\pm}$ \cite{AGMM:2010}.

Next, the lagrangian that yields the equations of motion (\ref{NVeq}) has to be found. There is a systematic prescription for doing this that is based on the soldering mechanism \cite{BAN:2002}. The basic lagrangians (\ref{Sumos}) have to be appropriately soldered or combined. It may be done in distinct ways. Consider first the possibility,
\begin{equation}
L=L_1-L_2=\frac{1}{2}(\dot{x}_1^2-\omega_1^2{x}_1^2)-\frac{1}{2}(\dot{x}_2^2-\omega_2^2{x}_2^2)
\label{Ldiff}
\end{equation}
Expressed in terms of the variables $z_{\pm}$ the lagrangian has the form,
\begin{equation}
L=\frac{1}{2}\dot{z}_{+}\dot{z}_{-}-\frac{1}{4}(\omega_1^2+\omega_2^2)z_{+}{z_-}-\frac{1}{8}(\omega_1^2-\omega_2^2)(z_{+}^2+z_{-}^2)
\label{Lagz}
\end{equation}
It is simple to check that the equations of motion (\ref{NVeq}) follow from the above lagrangian which therefore represents the PU model.

Some comments concerning (\ref{Lagz}) vis-a-vis the usual PU model (\ref{PUlag}) are useful. The new lagrangian has only usual (first order) derivatives. It is unconstrained and has two degrees of freedom $( z_{+},z_{-})$, exactly similar to the PU model\footnote{Observe that the standard method of converting the higher derivative PU model to a usual first order form by extending the space through a Lagrange multiplier converts it into a constrained system }. The hamiltonian analysis of (\ref{Lagz}) is therefore straightforward compared to (\ref{PUlag}) where Ostrogradsky's method or other similar approaches [\cite{MD:2005},\cite{BAGCHI:2013}, \cite{AGMM:2010}] have to be adopted.

The hamiltonian obtained from (\ref{Lagz}) by the usual Legendre transform has the form,
\begin{equation}
H=2p_{+}p_{-}+\frac{1}{4}(\omega_1^2+\omega_2^2)z_{+}{z_-}+\frac{1}{8}(\omega_1^2-\omega_2^2)(z_{+}^2+z_{-}^2)
\label{Hamz}
\end{equation}
where $p_{\pm}$ are the momenta conjugate to $z_{\pm}$,
\begin{equation}
p_{\pm}=\frac{\partial L}{\partial \dot{z}_{\pm}}=\frac{1}{2}\dot{z}_{\mp}
\end{equation}

It is simple to verify that the Hamilton's equation of motion obtained from (\ref{Hamz}) correctly yield (\ref{NVeq}). It is now possible to diagonalise the hamiltonian (\ref{Hamz}) by means of the canonical transformation,
\begin{equation}
z_{\pm}=x_1 \pm x_2,~~~~p_{\pm}=\frac{p_1 \pm p_2}{2}
\label{Diagz}
\end{equation}   
where ($x_1,p_1$) and ($x_2,p_2$) are the new canonical pairs. Note that the first transformation in (\ref{Diagz}) is identical to (\ref{NVar}). Moreover the above canonical transformation is applicable both for the nondegenerate $(\omega_1 \neq \omega_2)$ or degenerate $(\omega_1 = \omega_2)$ case. The standard canonical transformation (\ref{Diagct}) is applicable only for the nondegenerate case. The new hamiltonian is given by,
\begin{equation}
H= \frac{1}{2} (p_1^2 + {\omega}_1^2 x_1^2) - \frac{1}{2} (p_2^2 + {\omega}_2^2 x_2^2)
\label{HDIFF1}
\end{equation}
which reproduces the result (\ref{Hdiff}) obtained from the conventional Ostrogradski (or its first order version) method followed by the canonical transformation (\ref{Diagct}).

Let us now consider a different way of soldering the basic lagrangians (\ref{Sumos}). Instead of (\ref{Ldiff}), consider the sum of $L_1$ and $L_2$,
\begin{equation}
\tilde{L}=L_1+L_2=\frac{1}{2}(\dot{x}_1^2-\omega_1^2{x}_1^2)+\frac{1}{2}(\dot{x}_2^2-\omega_2^2{x}_2^2)
\end{equation}

In terms of the redefined variables $z_{\pm}$ (\ref{NVar}), the above lagrangian is given by,
\begin{equation}
\tilde{L}=\frac{1}{4}(\dot{z}_{+}^2+\dot{z}_{-}^2)-\frac{1}{8}(\omega_1^2+\omega_2^2)(z_{+}^2+z_{-}^2)-\frac{1}{4}(\omega_1^2-\omega_2^2)z_{+}z_{-}
\label{Ltildz}
\end{equation}

It is now easy to check that this lagrangian also reproduces the set of equations (\ref{NVeq}). Hence it represents the PU model. Moreover, it has the same nice features of the earlier lagrangian (\ref{Lagz}); i.e. has the correct degrees of freedom and characterises an unconstrained first order theory.

A usual Legendre transform immediately yields the hamiltonian corresponding to (\ref{Ltildz}). It is given by,
\begin{equation}
\tilde{H}={p_+}^2+{p_-}^2+\frac{1}{8}(\omega_1^2+\omega_2^2)(z_{+}^2+z_{-}^2)+\frac{1}{4}(\omega_1^2-\omega_2^2)z_{+}z_{-}
\label{Htildz}
\end{equation}
where as before, $p_{\pm}$ are the momenta conjugate to $z_{\pm}$,
\begin{equation}
p_{\pm}=\frac{\partial \tilde{L}}{\partial \dot{z}_{\pm}}=\frac{1}{2}\dot{z}_{\pm}
\end{equation}

As a consistency, one may check that the equations of motion (\ref{NVeq}) are reproduced from (\ref{Htildz}). The hamiltonian (\ref{Htildz}) is now diagonalised by means of the same canonical transformation (\ref{Diagz}). One obtains,
\begin{equation}
\tilde{H}= \frac{1}{2} (p_1^2 + {\omega}_1^2 x_1^2) + \frac{1}{2} (p_2^2 + {\omega}_2^2 x_2^2)
\label{Hco}
\end{equation}
Interestingly, the above expression turns out to be the sum of the hamiltonians of two decoupled harmonic oscillators. Obviously there is no problem in quantising such a system following the standard prescription. This enables a successful quantisation of the PU oscillator that is devoid of the boundedness or ghost problem.

It is now possible to understand the reason for obtaining the indefinite hamiltonian (\ref{Hdiff}) or (\ref{HDIFF1}), rather than the positive definite (\ref{Hco}), when using the original PU variables (\ref{PUlag}). The point is that whereas both (\ref{Lagz}) and (\ref{Ltildz}) yield the PU equation of motion (\ref{EOM}), only (\ref{Lagz}) leads to the original higher derivative lagrangian (\ref{PUlag}). This is shown by realising that, upto a total time derivative, (\ref{Lagz}) may be expressed as\footnote{An identical analysis goes through if the kinetic term is written as $-\frac{1}{2}{z}_{+}\ddot{z}_{-}$},
\begin{equation}
L=-\frac{1}{2}\ddot{z}_{+}{z}_{-}-\frac{1}{4}(\omega_1^2+\omega_2^2)z_{+}{z_-}-\frac{1}{8}(\omega_1^2-\omega_2^2)(z_{+}^2+z_{-}^2)
\label{Lagz1}
\end{equation}
Now $z_{-}$ becomes a nondynamical (auxilary) variable that may be eliminated, at least classically, by using the appropriate equation of motion (11). The result is 
\begin{equation}
L= \frac{1}{2(\omega_1^2 - \omega_2^2)} 
[{\ddot{z}^2}_{+} - ({\omega}_1^2 + {\omega}_2^2){\dot{z}^2}_{+} + \omega_1^2 \omega_2^2{z_{+}}^2]
\end{equation}
which reproduces, modulo an overall normalisation, the original PU lagrangian (\ref{PUlag}). Furthermore, the normalisation clearly shows the singular nature of the degenerate theory, ${\omega}_1 = {\omega}_2$. A similar treatment is not possible for (\ref{Ltildz}) where both $z_{+}$ and $z_{-}$ are dynamical variables.

The PU model, in its original higher derivative version, leads to the hamiltonian (\ref{Hdiff}) which is the genesis of all the (apparent) problems. The clue to our resolution lies in (\ref{Prodos}) which reveals that the PU model may be construed as a coupled set of usual harmonic oscillators. In order to unravel the precise meaning of this coupling, we started from the basic lagrangians (\ref{Sumos}) of the usual harmonic oscillators. Following the soldering mechanism we arrived at the soldered form of the lagrangian. This could be done in two different ways. In both cases the lagrangian turned out to be the usual (first order derivative) type with no constraints. Consequently the hamiltonian formulation was straightforward. In one case the hamiltonian turned out to be the difference of the hamiltonians of two harmonic oscillators thereby reproducing the standard result for the PU model. However, the other case was dramatically different. Instead of the difference, the result was found to be the sum of the hamiltonians of the harmonic oscillators which naturally resolves all the problems associated with the quantisation of the model. Observe that in getting at this result no imaginary scaling transformation, either at the classical or quantum level, was necessary. The quantum theory has a well defined classical limit and recourse to PT symmetry or a different norm than the usual one, as required in \cite{BM:2008}, are avoided.      

Our paper has highlighted the role of parametrization or choice of variables in analysing the Pais-Uhlenbeck (PU) oscillator. This is not surprising and occurs in various branches of physics. A proper choice of coordinates or variables may considerably simplify a problem whereas another choice would render it messy, if not downright misleading \cite{FPW}. Something similar occurs here. We have been successful in systematically obtaining a realization where all problems associated with the PU model are eliminated within the ambit of standard quantisation. It is a simple and conceptually clean approach which, contrary to some other approaches [\cite{MD:2005}, \cite{MOS:2011}], accommodates both the nondegenetrate $(\omega_1 \neq \omega_2)$ or degenerate $(\omega_1 = \omega_2)$ examples in a unified manner. We feel that our approach is both flexible and robust to discuss similar issues in the context of higher derivative gravity \cite{CLI:2012}. That the soldering mechanism was fruitfully employed to study several properties of the new massive gravity \cite{BAN:2012} further bolsters our hope. The present approach may also be applied to coupled oscillators with loss and gain \cite{BG} as will be discussed elsewhere.

\end{document}